\documentclass[12pt]{article}
\usepackage[dvips]{graphicx}
\usepackage{amsmath}
\usepackage{graphicx}
\usepackage{bm}
\usepackage{fancyhdr}
\usepackage{amssymb}
\begin{document}



\def\a{\alpha}
\def\b{\beta}
\def\d{\delta}
\def\e{\epsilon}
\def\g{\gamma}
\def\h{\mathfrak{h}}
\def\k{\kappa}
\def\l{\lambda}
\def\o{\omega}
\def\p{\wp}
\def\r{\rho}
\def\t{\tau}
\def\s{\sigma}
\def\z{\zeta}
\def\x{\xi}
\def\V={{{\bf\rm{V}}}}
 \def\A{{\cal{A}}}
 \def\B{{\cal{B}}}
 \def\C{{\cal{C}}}
 \def\D{{\cal{D}}}
\def\K{{\cal{K}}}
\def\O{\Omega}
\def\R{\bar{R}}
\def\T{{\cal{T}}}
\def\L{\Lambda}
\def\f{E_{\tau,\eta}(sl_2)}
\def\E{E_{\tau,\eta}(sl_n)}
\def\Zb{\mathbb{Z}}
\def\Cb{\mathbb{C}}

\def\R{\overline{R}}

\def\beq{\begin{equation}}
\def\eeq{\end{equation}}
\def\bea{\begin{eqnarray}}
\def\eea{\end{eqnarray}}
\def\ba{\begin{array}}
\def\ea{\end{array}}
\def\no{\nonumber}
\def\le{\langle}
\def\re{\rangle}
\def\lt{\left}
\def\rt{\right}

\def\sh{\sinh}
\def\ch{\cosh}
\def\tnh{\tanh}
\def\cth{\coth}

\newtheorem{Theorem}{Theorem}
\newtheorem{Definition}{Definition}
\newtheorem{Proposition}{Proposition}
\newtheorem{Lemma}{Lemma}
\newtheorem{Corollary}{Corollary}
\newcommand{\proof}[1]{{\bf Proof. }
        #1\begin{flushright}$\Box$\end{flushright}}

\baselineskip=20pt

\newfont{\elevenmib}{cmmib10 scaled\magstep1}
\newcommand{\preprint}{
   \begin{flushleft}
   \end{flushleft}\vspace{-1.3cm}
   \begin{flushright}\normalsize
   \end{flushright}}
\newcommand{\Title}[1]{{\baselineskip=26pt
   \begin{center} \Large \bf #1 \\ \ \\ \end{center}}}
\newcommand{\Author}{\begin{center}
   \large \bf
Yuan-Yuan Li${}^{a}$, Junpeng Cao${}^{a,b}$, Wen-Li Yang${}^{c,d}$,
Kangjie Shi${}^c$ and~Yupeng Wang${}^{a,b}\footnote{Corresponding
author: yupeng@iphy.ac.cn}$
 \end{center}}
\newcommand{\Address}{\begin{center}
     ${}^a$Beijing National Laboratory for Condensed Matter
           Physics, Institute of Physics, Chinese Academy of Sciences, Beijing
           100190, China\\
     ${}^b$Collaborative Innovation Center of Quantum Matter, Beijing,
     China\\
     ${}^c$Institute of Modern Physics, Northwest University,
     Xian 710069, China\\
     ${}^d$Beijing Center for Mathematics and Information Interdisciplinary Sciences, Beijing, 100048,  China
   \end{center}}
\newcommand{\Accepted}[1]{\begin{center}
   {\large \sf #1}\\ \vspace{1mm}{\small \sf Accepted for Publication}
   \end{center}}

\preprint \thispagestyle{empty}
\bigskip\bigskip\bigskip

\Title{Thermodynamic limit and surface energy of the XXZ spin chain
with arbitrary boundary fields} \Author

\Address \vspace{1cm}

\begin{abstract}

In two previous papers \cite{cysw2,cysw3}, the exact solutions of
the spin-$\frac12$ chains with arbitrary boundary fields were
constructed via the off-diagonal Bethe ansatz (ODBA). Here we
introduce a method to approach the thermodynamic limit of those
models. The key point is that at a sequence of degenerate points of
the crossing parameter $\eta=\eta_m$, the off-diagonal Bethe ansatz
equations (BAEs) can be reduced to the conventional ones. This
allows us to extrapolate the formulae derived from the reduced BAEs
to arbitrary $\eta$ case with $O(N^{-2})$ corrections in the
thermodynamic limit $N\to\infty$. As an example, the surface energy
of the $XXZ$ spin chain model with arbitrary boundary magnetic
fields is derived exactly. This approach can be generalized to all
the ODBA solvable models.

\vspace{1truecm} \noindent {\it PACS:} 75.10.Pq, 02.30.Ik, 71.10.Pm


\end{abstract}

\newpage






\section{Introduction}
The integrable models have played very important roles in
statistical physics \cite{bax}, quantum field theory\cite{fad} and
low-dimensional condensed matter physics \cite{kor1,kor2}. In the
recent years, new applications have been found on cold atom systems
and AdS/CFT correspondence. For examples, the Lieb-liniger model
\cite{lieb1, lieb2}, Yang model \cite{yang} and the one-dimensional
Hubbard model \cite{lieb-wu} have provided important benchmarks for
the one-dimensional cold atom systems and even fitted experimental
data with incredibly high accuracy \cite{cold}. On the other hand,
the anomalous dimensions of single-trace operators of ${\cal{N}}=4$
super-symmetric Yang-Mills (SYM) field theory can be given by the
eigenvalues of certain closed integrable spin chains
\cite{Min03,Bei12} while the anomalous dimensions of the
determinant-like operators of ${\cal{N}}=4$ SYM \cite{Ber05,Hof07}
can be mapped to the eigenvalue problem of certain open integrable
spin chains with boundary fields \cite{Mur08,Nep11,Bei12}. By
AdS/CFT correspondence the boundaries correspond to open strings
attached to maximal giant gravitons \cite{McG00,Hof07}. Sometimes
those boundaries may even break the $U(1)$ symmetry.

Indeed, among the family of quantum integrable models, there exists
a large class of models which do not possess $U(1)$ symmetry and
make the conventional Bethe ansatz methods such as coordinate Bethe
ansatz \cite{co1,co2}, algebraic Bethe ansatz \cite{ab1,ab2} and
$T-Q$ relation \cite{tq1,tq2} quite hard to be used because of
lacking a proper reference state. Some famous examples are the $XYZ$
spin chain with odd number of sites \cite{tak}, the anisotropic spin
torus \cite{bat,cysw1} and the quantum spin chains with non-diagonal
boundary fields \cite{cysw2,cysw3,nepo3,nepo31}. Those models have
been realized also possessing important applications in
non-equilibrium statistical physics (e.g., stochastic processes
\cite{stoc1,Gie05,stoc2}), in condensed matter physics (e.g., a
Josephson junction embedded in a Luttinger liquid \cite{jose},
spin-orbit coupling systems, one-dimensional cold atoms coupled with
a BEC reservoir etc.) and in high energy physics (e.g., open strings
and coupled D-Branes). Very recently, a systematic method for
solving the integrable models without $U(1)$ symmetry, i.e., the
so-called off-diagonal Bethe ansatz (ODBA) method was proposed
\cite{cysw1,cysw2,cysw3} and several long-standing models were
solved exactly \cite{cysw1,cysw2,cysw3,lcysw,nepo1,cysw4,cysw5}.
However, an important issue about this kind of models, i.e., the
thermodynamic limit, is still open. The difficulty to approach the
thermodynamic limit of those models lies in that there is an
off-diagonal term (or inhomogeneous term) in the Bethe ansatz
equations (BAEs), which makes the distributions of the Bethe roots
quite opaque.

In this paper, we propose that the thermodynamic limit of the ODBA
solvable models for arbitrary crossing parameter $\eta$ can be
derived from those at a sequence of degenerate points $\eta=\eta_m$
up to the order $O(N^{-2})$. At these special points, the ODBA
equations are reduced to the usual BAEs which allow us to use the
usual tools to derive the thermodynamic quantities. As
$\eta_{m+1}-\eta_m=2i\pi/N$, those degenerate points become dense in
the thermodynamic limit $N\to\infty$. In the following text, we take
the $XXZ$ spin chain model with arbitrary boundary fields as an
example to elucidate how the method works.

The paper is organized as follows: In the next section, the
Hamiltonian and the associated ODBA equations are introduced.
Sec.III is attributed to the calculation of the surface energy at
the degenerate points $\eta=\eta_m$. The analysis about arbitrary
$\eta$ case is given in Sec.IV. Concluding remarks and discussions
are given in Sec.V.

\section{The model and its ODBA solutions}

Let us consider a typical ODBA solvable model, i.e., the XXZ spin
chain with arbitrary boundary fields. The Hamiltonian reads
\begin{eqnarray}
H=\sum_{j=1}^{N-1}\left[\sigma_{j}^{x}\sigma_{j+1}^{x}+\sigma_{j}^{y}
\sigma_{j+1}^{y}+\ch\eta\sigma_{j}^{z} \sigma_{j+1}^{z}\right]
+{\vec h}_-\cdot\vec \sigma_1+\vec h_+\cdot\vec\sigma_N,
\end{eqnarray}
where $\sigma_j^\alpha$ ($\alpha=x,y,z$) are the Pauli matrices as
usual and $\vec h_{\pm}=(h_{\pm}^x, h_{\pm}^y, h_{\pm}^z)$ are the
boundary magnetic fields. For convenience, we adopt the notations in
Ref. \cite{cysw3} to parameterize the boundary fields as
\begin{eqnarray}
h_{\pm}^x=\frac{\sh\eta\ch\theta_{\pm}}{\sh\alpha_{\pm}\ch\beta_{\pm}},\quad
h_{\pm}^y=\frac{i\sh\eta\sh\theta_{\pm}}{\sh\alpha_{\pm}\ch\beta_{\pm}},
\quad h_{\pm}^z=\mp\sh\eta\cth\alpha_{\pm}\tnh\beta_{\pm}.
\end{eqnarray}
The eigenvalues of the Hamiltonian thus read
\begin{eqnarray}
&&E=-\sh\eta[\cth(\alpha_-)+\tnh(\beta_-)+\cth(\alpha_+)\hspace{1.2cm}
\nonumber
\\ && \qquad +\tnh(\beta_+)+2
\sum_{j=1}^{M}\cth(\mu_j+\eta) -(N-1)\cth\eta],\label{2Spectrumtt}
\end{eqnarray}
where the Bethe roots $\mu_j$ are determined by the ODBA equations
\begin{eqnarray}
&&\frac{\bar{c}\sh(2\mu_j+\eta)\sh(2\mu_j+2\eta)} {2
\sh(\mu_j+\alpha_-+\eta)\ch(\mu_j+\beta_-+\eta)} \frac{\sh^n\mu_j\sh^{M+N}(\mu_j+\eta)} {\sh(\mu_j+\alpha_++\eta)\ch(\mu_j+\beta_++\eta) }\nonumber\\
&&=
\prod_{l=1}^M\sh(\mu_j+\mu_l+\eta)\sh(\mu_j+\mu_l+2\eta),\label{BAE}
\end{eqnarray}
$j=1,\ldots,M$ and
\begin{eqnarray}
\bar{c}=\ch\Big[(N+2n+1)\eta+\alpha_-+\beta_-+\alpha_++\beta_+ +
2\sum_{j=1}^M\mu_j\Big] -\ch(\theta_--\theta_+), \label{c-para-1}
\end{eqnarray} with $n$ a non-negative even (odd) integer \footnote{In reference \cite{cysw3}, $n=0$ for even $N$
and $n=1$ for odd $N$ were adopted. The $T-Q$ relation with
arbitrary $n$ was considered in reference \cite{cysw2}.} for even
(odd) $N$ and $M=N+n$. Interestingly, when the boundary parameters
and the crossing parameter $\eta$ satisfy the following constraint
condition \cite{cysw3,cslw}
\begin{eqnarray}
(2M_1-N+1)\eta+\alpha_-+\beta_-+\alpha_++\beta_+
\pm(\theta_--\theta_+)=2\pi im,
\end{eqnarray}
there does exist a solution to (\ref{BAE})-(\ref{c-para-1}) such
that the parameter $\bar c=0$ and hence the Bethe roots are
classified into two types of pairs
\begin{eqnarray}
(\mu_l,-\mu_l-\eta),\quad\quad (\mu_l,-\mu_l-2\eta),\nonumber
\end{eqnarray}
with $M_1$ the number of the first pairs and $m$ an arbitrary
integer.

Let us focus on the gapless region, i.e., imaginary $\eta$ and
$\theta_\pm$ case. Without losing generality, we put $\alpha_\pm$
imaginary and $\beta_\pm$ real to ensure the boundary fields being
real. Let us examine the solutions at the degenerate points
$\eta=\eta_m$ (corresponding to the case of $\bar{c}=0$ and $M_1=N$)
and $\beta_\pm=\pm\beta$,
\begin{eqnarray}
\eta_m=-\frac{\alpha_-+\alpha_+\pm(\theta_--\theta_+)+2\pi
im}{N+1}.\label{Degenerate-P}
\end{eqnarray}
For convenience, let us take $\lambda_j=\mu_j+\frac{\eta}2$,
$ia_{\pm}=\alpha_\pm+\frac{\eta}2$, $\eta=i\theta$, with
$a_\pm,\theta\in(0,{\pi})$. With these parameters, the reduced BAEs
for $\eta=\eta_m$ become\footnote{The reduced BAEs were derived from
the regularity of the reduced $\L(u)$ in\cite{cysw3}. See
also\cite{cslw}}
\begin{eqnarray}
&&\left[\frac{\sh(\lambda_j-i\frac{\theta}2)}{\sh(\lambda_j+i\frac{\theta}2)}\right]^{2N}
\frac{\sh(2\lambda_j-i\theta)}{\sh(2\lambda_j+i\theta)}\frac{\sh(\lambda_j+ia_+)}{\sh(\lambda_j-ia_+)}
\nonumber \\
&&\qquad \times
\frac{\sh(\lambda_j+ia_-)}{\sh(\lambda_j-ia_-)}\frac{\ch(\lambda_j+\beta+i\frac{\theta}2)}{\ch(\lambda_j+\beta-i\frac{\theta}2)}
\frac{\ch(\lambda_j-\beta+i\frac{\theta}2)}{\ch(\lambda_j-\beta-i\frac{\theta}2)}
\nonumber\\
&&=-\prod_{l=1}^N\frac{\sh(\lambda_j-\lambda_l-i\theta)\sh(\lambda_j+\lambda_l-i\theta)}
{\sh(\lambda_j-\lambda_l+i\theta)\sh(\lambda_j+\lambda_l+i\theta)},\label{BAE-2}
\end{eqnarray}
where $j=1,\ldots, N$. The above reduced BAEs were firstly observed
in \cite{cslw}. The corresponding eigenenergy is given by
\begin{eqnarray}
&& E=-
\sum_{j=1}^{N}\frac{4\sin^2\theta}{\ch(2\lambda_j)-\cos\theta}
-\sin\theta[\cot(a_+-\theta/2)\nonumber \\ [6pt]&&\qquad
+\cot(a_--\theta/2)]+(N-1)\cos\theta.
\end{eqnarray}

We confirm that for $\eta=\eta_m$, the reduced BAEs (\ref{BAE-2})
give a complete set of solutions as verified numerically
\cite{nepo}. Here, we have checked this statement numerically for
small $N$. The numerical solutions of (\ref{BAE-2}) for $N=3,4$ with
randomly chosen boundary parameters are shown in Table 1 $\&$ 2,
respectively. The eigenvalues $E$ of the Hamiltonian shown in the
Tables are exactly the same as those from exact diagonalization.
\\
\begin{table}
\caption{\label{N3} The numerical solutions of (\ref{BAE-2}) for
$N=3$ with the parameters $\eta=-i$, $\alpha_+=2i$, $\alpha_-=3i$,
$\beta_+=1$, $\beta_-=-1$, $\theta_+=2i$, $\theta_-=i$. $el$
indicates the number of the energy levels}
\begin{tabular}{ ccc|c|c} \hline \hline
$\mu_1$ & $\mu_2$ & $\mu_3$ & $E$ & $el$ \\ \hline
$-1.48510+0.67075i$ & $-1.48510+2.47085i$ & $0.36994-0.00000i$ & $-9.10664$ & $1$ \\
$-0.63430-1.57080i$ & $-0.38556+0.50089i$ & $-0.38556-0.50089i$ & $-5.80407$ & $2$ \\
$-1.11069+1.00247i$ & $-1.11069+2.13912i$ & $-1.10123-0.00000i$ & $-5.30177$ & $3$ \\
$-1.68396-0.65365i$ & $0.59457+1.57080i$ & $1.68396-0.65365i$ & $-4.08354$ & $4$ \\
$-0.51260-1.57080i$ & $-0.38055+0.00000i$ & $-0.00000+0.64158i$ & $3.46000$ & $5$ \\
$-1.56515-0.66501i$ & $-0.00000+0.64158i$ & $1.56515-0.66501i$ & $5.73191$ & $6$ \\
$-0.00000+0.64159i$ & $0.25391-1.57080i$ & $1.09544+0.00000i$ & $6.81205$ & $7$ \\
$-0.94157+1.57080i$ & $-0.00000+0.64159i$ & $0.20977+1.57080i$ & $8.29206$ & $8$ \\
\hline\hline \end{tabular}
\end{table}

{\tiny
\begin{table}
\caption{\label{N4} The numerical solutions of (\ref{BAE-2}) for
$N=4$ with the parameters $
\theta=1,a_+=2.5,a_-=1.5,\beta=1,\theta_-=3i,\theta_+=-5i,m=0$.}
\footnotesize
\begin{tabular}{ cccc|c|c} \hline \hline
$\lambda_1$ & $\lambda_2$ & $\lambda_3$ & $\lambda_4$ & $E$ & $el$ \\
\hline
$-1.66762-0.00000i$ & $-1.21632+1.57080i$ & $-0.55018+0.00000i$ & $-0.21316+0.00000i$ & $-5.93342$ & $1$ \\
$-0.92025-0.00000i$ & $-0.91893+0.99518i$ & $0.20501-3.14159i$ & $0.91893+0.99518i$ & $-3.85243$ & $2$ \\
$-0.94690-2.65582i$ & $-0.68831-1.57080i$ & $-0.19733-0.00000i$ & $0.94690+3.62737i$ & $-3.58123$ & $3$ \\
$-0.46931-1.57080i$ & $-0.17931-0.00000i$ & $1.43232+1.57080i$ & $1.90426-0.00000i$ & $-2.35148$ & $4$ \\
$-0.91410+0.00000i$ & $-0.91160-0.99287i$ & $-0.91160+0.99287i$ & $-0.50119+0.00000i$ & $-1.47531$ & $5$ \\
$-0.87562-3.62850i$ & $-0.87562+0.48690i$ & $-0.57755-1.57080i$ & $-0.45893-0.00000i$ & $-1.36888$ & $6$ \\
$0.39497-3.14159i$ & $0.39554+1.57080i$ & $1.40894+1.57080i$ & $1.87940+0.00000i$ & $-0.43122$ & $7$ \\
$-1.77682+0.00000i$ & $-0.48316+0.50058i$ & $0.48316+0.50058i$ & $1.32868-1.57080i$ & $0.08195$ & $8$ \\
$-1.40329+0.00000i$ & $-0.89579-2.13792i$ & $0.89236+0.00000i$ & $0.89579+1.00367i$ & $0.98414$ & $9$ \\
$-0.55149+0.49964i$ & $-0.38646-1.57080i$ & $0.55149+0.49964i$ & $1.39986+0.00000i$ & $1.15594$ & $10$ \\
$-1.80115-0.00000i$ & $-1.32916+1.57080i$ & $-0.72884-0.00000i$ & $-0.28785+1.57080i$ & $1.68692$ & $11$ \\
$-0.92875-1.57080i$ & $-0.90786+0.99956i$ & $-0.90786+2.14204i$ & $-0.90693+0.00000i$ & $2.08877$ & $12$ \\
$-0.86225+1.57080i$ & $-0.46738-2.64162i$ & $0.33674-1.57080i$ & $0.46738+0.49997i$ & $2.26194$ & $13$ \\
$-0.93579+0.99864i$ & $-0.93536-0.00000i$ & $0.23087+1.57080i$ & $0.93579-2.14295i$ & $3.06940$ & $14$ \\
$0.22280+1.57080i$ & $0.92385-1.57080i$ & $1.13160-0.48933i$ & $1.13160+0.48933i$ & $3.33186$ & $15$ \\
$-1.56849-1.57080i$ & $-0.75387+1.57080i$ & $0.19162+1.57080i$ & $2.05390+0.00000i$ & $4.33306$ & $16$ \\
\hline\hline \end{tabular} \\
\end{table}
}

\section{The surface energy for $\eta=\eta_m$}

Let us consider the ground state energy at the degenerate crossing
parameter points given by (\ref{Degenerate-P}). Since a real
$\lambda_j$ contributes negative energy, the Bethe roots should fill
the real axis as long as possible. However, in the thermodynamic limit, the maximum number of
Bethe roots accommodated by the real axis is only about $N/2$, some of the
roots must be repelled to the complex plane and form a string
\cite{takahashi}. Suppose there is a $k$ string\footnote{Another
type of strings may exist in this model. Different choice of the
bulk string does not affect the surface energy as the string's
contribution to the ground state energy is zero in the thermodynamic
limit. For rational $\pi/\eta$, there is a constraint for $k$. Here
we consider the case of $\pi/\eta_m$ away from those special values.
In fact we can always take $N$ a prime number to ensure the possible
$k$ being large enough.  For detail, see \cite{takahashi}.} in the
ground state configuration with
\begin{eqnarray}
\lambda_l^s=\lambda^r+i\frac{\theta}{2}(k+1-2l)+O(e^{-\delta N}),\;
l=1,\ldots,k,\label{solution}
\end{eqnarray}
where $\lambda^r$ is the position of the string on the real axis and
$\delta$ is a positive number to account for the small deviation.
Substituting (\ref{solution}) into (\ref{BAE-2}) and omitting the
exponentially small corrections we obtain
\begin{eqnarray}
&&\left[\frac{\sh(\lambda_j-i\frac{\theta}2)}{\sh(\lambda_j+i\frac{\theta}2)}\right]^{2N}
\frac{\sh(2\lambda_j-i\theta)}
{\sh(2\lambda_j+i\theta)}\frac{\sh(\lambda_j+ia_+)}{\sh(\lambda_j-ia_+)}
\nonumber\\
&&\qquad \times
\frac{\sh(\lambda_j+ia_-)}{\sh(\lambda_j-ia_-)}\frac{\ch(\lambda_j+\beta+i\frac{\theta}2)}{\ch(\lambda_j+\beta-i\frac{\theta}2)}
\frac{\ch(\lambda_j-\beta+i\frac{\theta}2)}{\ch(\lambda_j-\beta-i\frac{\theta}2)}
\nonumber\\
&&=-\prod_{l=1}^{N-k}\frac{\sh(\lambda_j-\lambda_l-i\theta)\sh(\lambda_j+\lambda_l-i\theta)}
{\sh(\lambda_j-\lambda_l+i\theta)\sh(\lambda_j+\lambda_l+i\theta)}\label{BAE-ns} \\
&&\qquad \times \frac{\sh(\lambda_j+\lambda^r-i
\frac{\theta}{2}(k+1))\sh(\lambda_j+\lambda^r-i\frac{\theta}{2}(k-1))}
{\sh(\lambda_j+\lambda^r+i\frac{\theta}{2}(k+1))
\sh(\lambda_j+\lambda^r+i\frac{\theta}{2}(k-1))} \nonumber \\
&&\qquad
\times\frac{\sh(\lambda_j-\lambda^r-i\frac{\theta}{2}(k+1))\sh(\lambda_j-\lambda^r-i\frac{\theta}{2}(k-1))}
{\sh(\lambda_j-\lambda^r+i\frac{\theta}{2}(k+1))\sh(\lambda_j-\lambda^r+i\frac{\theta}{2}(k-1))},\nonumber
\end{eqnarray}
where $j=1,\ldots, N-k$.

We consider the $a_\pm\in(\frac\pi2,{\pi})$ case. Taking the
logarithm of (\ref{BAE-ns}) we have
\begin{eqnarray}
&&\phi_1(\lambda_j)+\frac{1}{2N}[\phi_2(2\lambda_j)-\phi_{2a_+/\theta}(\lambda_j)-\phi_{2a_-/\theta}(\lambda_j)
+B(\lambda_j+\beta)+B(\lambda_j-\beta)
\nonumber\\[6pt]
&&\qquad -\pi-\phi_{k+1}(\lambda_j-\lambda^r) -\phi_{k-1}(\lambda_j-\lambda^r)-\phi_{k+1}(\lambda_j+\lambda^r)-\phi_{k-1}(\lambda_j+\lambda^r)] \nonumber\\[6pt]
&&=2\pi\frac{I_j}{2N}+\frac1{2N}\sum_{l=1}^{N-k}[\phi_2(\lambda_j-\lambda_l)+\phi_2(\lambda_j+\lambda_l)],\label{BAE-log}
\end{eqnarray}
where $I_j$ is an integer and
\begin{eqnarray}
&&\phi_m(\lambda_j)=-i\ln\frac{\sh(\lambda_j-i\frac{m\theta}2)}{\sh(\lambda_j+i\frac{m\theta}2)}\nonumber\\
&&B(\lambda_j)=-i\ln\frac{\ch(\lambda_j+i\frac{\theta}{2})}{\ch(\lambda_j-i\frac{\theta}{2})}.
\end{eqnarray}
For convenience, let us put $\lambda_l=-\lambda_{-l}$ and define the
counting function $Z(\lambda)$ as
\begin{eqnarray}
&& Z(\lambda)=\frac{1}{2\pi}\Bigg\{\phi_1(\lambda)+\frac
{1}{2N}\Bigg
[\phi_2(2\lambda)-\phi_{2a_+/\theta}(\lambda)-\phi_{2a_-/\theta}(\lambda)+B(\lambda+\beta)
\nonumber
\\ [6pt]&&\qquad\qquad +B(\lambda-\beta)
-\phi_{k+1}(\lambda-\lambda^r) -\phi_{k-1}(\lambda-\lambda^r)-\phi_{k+1}(\lambda+\lambda^r) \nonumber\\[6pt]
&&\qquad\qquad
-\phi_{k-1}(\lambda+\lambda^r)-\pi-\sum_{l=1}^{N-k}[\phi_2(\lambda-\lambda_l)
+\phi_2(\lambda+\lambda_l)]\Bigg]\Bigg\}.
\end{eqnarray}
Obviously, $Z(\lambda_j)=I_j/2N$ coincides with Eq.(\ref{BAE-log}).
In the thermodynamic limit $N\rightarrow\infty$, the density of the
real roots $\rho(\lambda)$ is
\begin{eqnarray}
&&\rho(\lambda)=\frac{d Z(\lambda)}{d \lambda}-\frac{1}{2N}\delta(\lambda)\nonumber\\[6pt]
&&\qquad\;
=a_1(\lambda)+\frac{1}{2N}[2a_2(2\lambda)-a_{2a_+/\theta}(\lambda)
-a_{2a_-/\theta}(\lambda) +b(\lambda+\beta)+b(\lambda-\beta)
\nonumber \\ [6pt]&& \qquad\qquad -a_{k+1}(\lambda-\lambda^r)
-a_{k-1}(\lambda-\lambda^r)-a_{k+1}(\lambda+\lambda^r)-a_{k-1}(\lambda+\lambda^r)\nonumber\\[6pt]
&&\qquad\qquad -\delta(\lambda)] -\int_{-\infty}^\infty
a_2(\lambda-\nu)\rho(\nu)d\nu,\label{density}
\end{eqnarray}
with
\begin{eqnarray}
&&a_m(\lambda)=\frac{1}{2\pi}\frac{d\phi_m(\lambda)}{d\lambda}=\frac{1}{\pi}\frac{\sin
m\theta}{\ch 2\lambda-\cos m\theta},\label{am}\\[6pt]
&&
b(\lambda)=\frac{1}{2\pi}\frac{dB(\lambda)}{d\lambda}=\frac{1}{\pi}\frac{\sin\theta}{\ch(2\lambda)+\cos\theta}.
\end{eqnarray}
where the $\delta(\lambda)$ term accounts for the hole at
$\lambda=0$ which is a solution of the BAEs but can never be
occupied in any case. With the Fourier transformations
\begin{eqnarray}
\hat f(\omega)=\int_{-\infty}^\infty
f(\lambda)e^{i\omega\lambda}d\lambda,\nonumber
\end{eqnarray}
we obtain
\begin{eqnarray}
\hat\rho(\omega)=\hat\rho_0(\omega)+\hat\rho_b(\omega),
\end{eqnarray}
where
\begin{eqnarray}
&&\hat\rho_0(\omega)=\frac{\hat a_1(\omega)}{1+\hat
a_2(\omega)},\label{rho-0}\\[6pt]
&&\hat\rho_b(\omega)=\frac{1}{2 N[1+\hat a_2(\omega)]}\Big\{\hat
a_2({\omega\over2})-\hat a_{2a_+/\theta}(\omega)  -\hat
a_{2a_-/\theta}(\omega) +2\cos(\beta\omega)\hat
b(\omega)\nonumber\\[6pt] && \qquad\qquad -2\cos(\lambda^r\omega)\big[\hat
a_{k+1}(\omega)+\hat a_{k-1}(\omega)\big]-1\Big\},\\[6pt] &&\hat
a_m(\omega)=\frac{\sh(\pi\omega/2-\delta_m\pi\omega)}{\sh(\pi\omega/2)},
\quad  \hat b(\omega)=\frac{\sh(\theta\omega/2)}{\sh(\pi\omega/2)},
\end{eqnarray}
with
$\delta_m\equiv\frac{m\theta}{2\pi}-\lfloor\frac{m\theta}{2\pi}\rfloor$
denoting the fraction part of $\frac{m\theta}{2\pi}$. For
$\rho(\lambda)$ is the density of the real roots and $M_1=N$, the
following equation must hold
\begin{eqnarray}
N\int_{-\infty}^\infty\rho(\lambda)d\lambda+k=N,
\end{eqnarray}
which gives the length of the string $k$,
\begin{eqnarray}
k=\frac{N}{2}-\frac{a_++a_-+2\pi(\delta_{k+1}+\delta_{k-1})-3\pi}{2(\pi-\theta)}.
\end{eqnarray}
Obviously, $k$ has the order of $N/2$.

In the ground state, $\lambda_r\to\infty$ to minimize the energy.
The ground state energy in the thermodynamic limit can be easily
derived as
\begin{eqnarray}
&&E=-4\pi N\sin\theta\int_{-\infty}^\infty
a_1(\lambda)\rho(\lambda)d\lambda- \sin\theta [4\pi a_k(\lambda^r)
\nonumber \\ && \; \qquad +\cot(a_+-\theta/2)+\cot(a_--\theta/2)
-(N-1)\cot\theta ]\nonumber\\[6pt]
&&\quad \;=Ne_0+e_b,
\end{eqnarray}
and
\begin{eqnarray}
&&e_0=-\int_{-\infty}^\infty\frac{2\sin\theta\sh^2(\pi\omega/2-\theta\omega/2)}
{\sh(\pi\omega/2)[\sh(\pi\omega/2)+\sh(\pi\omega/2-\theta\omega)]}d\omega +\cos\theta,\\[6pt]
&&e_b=e_b^0+I_1(a_+)+I_1(a_-)+2I_2(\beta)\label{e1}
\end{eqnarray}
with $e_0$ the ground state energy density of the periodic chain and
$e_b$ the surface energy, where
\begin{eqnarray}
&&e_b^0=-\sin\theta\int_{-\infty}^\infty\frac{\hat
a_1(\omega)}{1+\hat a_2(\omega)}[\hat
a_2(\omega/2)-1]d\omega-\cos\theta,\nonumber\\[6pt]
&&I_1(\alpha)=\sin\theta\int_{-\infty}^\infty\frac{\hat
a_1(\omega)}{1+\hat a_2(\omega)}\hat
a_{2\alpha/\theta}(\omega)d\omega  -\sin\theta\cot(\alpha-\theta/2),\nonumber\\[6pt]
&&I_2(\beta)=-\sin\theta\int_{-\infty}^\infty\frac{\hat
a_1(\omega)}{1+\hat a_2(\omega)}\cos(\beta\omega)\hat
b(\omega)d\omega.\label{energy}
\end{eqnarray}
Some remarks are in order: (1)The extra string in the ground state
configuration contributes nothing to the energy in the thermodynamic
limit. However, for a finite $N$, the string may induce
exponentially small corrections. (2)Above we considered only the
parameter region $a_\pm\in(\pi/2,\pi)$. For the boundary parameters
out of this region, stable boundary bound states exist in the ground
state \cite{ham,batch,sal,surf}. However, the energy is indeed a smooth
function about the boundary parameters as demonstrated in the
diagonal boundary field case \cite{surf,Murgan}. (3)An interesting
fact is that the contributions of $a_+,a_-,\beta$ to the energy are
completely separated and the surface energy does not depend on
$\theta_\pm$ at all (same effect was also obtained in \cite{ahn} where the surface energy and the finite size correction were derived for some constraint boundary parameters), which indicate that the two boundary fields
behave independently in the thermodynamic limit. Similar phenomenon
often occurs in the dilute impurity systems. In such a sense, we may
adjust $\theta_\pm$ to match $\bar c=0$ for arbitrary $\eta$ and
non-negative integer $M_1$ without affecting the thermodynamic
quantities up to the order of $O(N^{-1})$.  We note the surface
energy does depend on the relative directions of the boundary fields
to the $z$-axis because of the anisotropy of the bulk. (4)In the
above calculations, we put the integral limits to infinity which is
reasonable to the surface energy. To account for the finite size
corrections of order $1/N$ (Casimir effect or central charge term),
one should keep a finite cutoff for the integrals. Calculations can
also be performed by the standard finite size correction and
Wiener-Hopf methods \cite{yang2,dev,ham,batch}. The correlations
between the two boundaries exist in this order \cite{batch,ahn}. (5)The
thermodynamic equations at the degenerate points $\eta=\eta_m$ can
also be derived by following the standard method \cite{takahashi}.
(6)When $\beta=0$, the boundary magnetic fields lie in the $x-y$
plane. Taking the limit $\eta\to0$ of Eq.(\ref{e1}) we obtain the
surface energy of the $XXX$ spin chain with arbitrary boundary
fields, which obviously does not depend on the angles $\theta_\pm$.
The $\theta_\pm$-dependence of the ground state energy only occurs
in the order of $1/N$ as verified by the numerical simulations
\cite{jccysw,Nep14}.

Now let us turn to arbitrary $\beta_\pm$ case. In this case, the
degenerate points of $\eta$ takes complex values and the above
derivations are invalid. However, we can deduce the surface energy
with the following argument. In principle, for $N\to\infty$ the
surface energy takes the form
\begin{eqnarray}
{\epsilon}_b=\epsilon_b^0+\bar\epsilon_b(\alpha_+,\beta_+,\theta_+)+\bar\epsilon_b(\alpha_-,\beta_-,\theta_-),\label{e-2}
\end{eqnarray}
because the two boundaries decouple completely as long as the bulk
is not long-range ordered. Here the second and the third terms are
the contributions of the boundary fields. For arbitrary real
$\beta_\pm$, suppose
\begin{eqnarray}
\bar\epsilon_b(\alpha_\pm,\beta_\pm,\theta_\pm)=I_1(a_\pm)+\bar
I(a_\pm,\beta_\pm,\theta_\pm).\label{e-3}
\end{eqnarray}
When $\beta_\pm=\pm\beta$, from Eqs.(\ref{e1}), (\ref{e-2}),
(\ref{e-3}) we have
\begin{eqnarray}
\bar I(a_+,\beta,\theta_+)+\bar I(a_-,-\beta,\theta_-)=2I_2(\beta),
\end{eqnarray}
which indicates that $\bar I(\alpha_\pm,\beta_\pm,\theta_\pm)$ does
not depend on $\alpha_\pm$ and $\theta_\pm$. In addition, for
$\alpha_-=i\pi/2$, the boundary field is an even function of
$\beta_-$. Since $\bar I(\alpha_-,\beta_-,\theta_-)$ is independent
of $\alpha_-,\theta_-$, it must be an even function of $\beta_-$.
The same conclusion holds for $\beta_+$. Therefore we conclude that
\begin{eqnarray}
{\epsilon}_b=\epsilon_b^0+I_1(a_+)+I_1(a_-)+I_2(\beta_+)+I_2(\beta_-).
\end{eqnarray}
The above formula is valid for arbitrary boundary fields and $\eta$
in the thermodynamic limit $N\to\infty$ since $\eta_m$ become dense.

\section{Physical quantities for large $N$ and generic $\eta$}

With the reduced BAEs at the degenerate $\eta$ points, most of the
physical quantities as functions of $\eta_m$ can be derived up to
the order of $1/N$ with the conventional methods, i.e.,
\begin{eqnarray}
\hspace{-2mm}F(\eta_m)=Nf_0(\eta_m)+f_1(\mu_m)+\frac1Nf_2(\eta_m)+O(N^{-2}).
\end{eqnarray}
Let us treat $f_n(\eta)$ ($n=0,1,2$) as known functions. For a
generic $i\eta_{m}\ge i\eta\ge i\eta_{m+1}$, we suppose that the
corresponding quantities are $\bar f_n(\eta)$ which are initially
unknown functions. We suppose further both $f_n(\eta)$ and $\bar
f_n(\eta)$ are smooth functions about $\eta$. Obviously,
\begin{eqnarray}
\bar f_n(\eta_m)=f_n(\eta_m),
\end{eqnarray}
and $\bar f_0(\eta)=f_0(\eta)$ because $f_0$ is boundary-field
independent and is the same calculated from the corresponding
periodic system. Let us make the following Taylor expansions around
$\eta_m$ and $\eta_{m+1}$ ($n=1,2$)
\begin{eqnarray}
&&\bar f_n(\eta)=\bar f_n(\eta_m)+\bar
f_n'(\eta_m)\bar\delta_1+O(N^{-2})\nonumber\\[6pt]
&&\qquad\;\;=f_n(\eta_m)+\bar f_n'(\eta_m)\bar\delta_1+O(N^{-2})\nonumber\\[6pt]
&&\qquad\;\;=f_n(\eta_{m+1})+\bar
f_n'(\eta_{m+1})\bar\delta_2+O(N^{-2}),
\end{eqnarray}
with $\bar\delta_1=\eta-\eta_m$ and
$\bar\delta_2=\bar\delta_1-\frac{2i\pi}{N}$. Notice that
\begin{eqnarray}
&&f_n(\eta_{m+1})=f_n(\eta_m)+f_n'(\eta_m)\frac{2i\pi}{N}+O(N^{-2}),\nonumber\\[6pt]
&&f_n'(\eta_{m+1})=f_n'(\eta_m)+O(N^{-1}),\nonumber
\end{eqnarray}
we readily have
\begin{eqnarray}
\bar f_n'(\eta_m)=f_n'(\eta_m)+O(N^{-1}),\nonumber
\end{eqnarray}
and
\begin{eqnarray}
&&\bar f_n(\eta)=f_n(\eta_m)+f_n'(\eta_m)\bar\delta_1+O(N^{-2})\nonumber\\[6pt]
&&\qquad\;\;=f_n(\eta)+O(N^{-2}),
\end{eqnarray}
which means that the unknown function $\bar f_n(\eta)$ can be
replaced by the known function $f_n(\eta)$ up to the order of
$O(N^{-2})$.

\section{Concluding remarks}

In conclusion, a systematic method is proposed for approaching the
thermodynamic limit of the ODBA solvable models with the open XXZ
spin chain as an example. The central idea of this method lies in
that at a sequence of degenerate crossing parameter points, the ODBA
equations can be reduced to the conventional BAEs, which allows us
to derive the thermodynamic quantities with the well developed
methods. We remark that there are no degenerate points for the
isotropic Heisenberg spin chain model \cite{cysw2}, the $XXZ$ model
for real $\eta$ and the $XXZ$ spin torus \cite{cysw1}. However, the
thermodynamic quantities can be observed from their anisotropic
correspondences. For the Heisenberg chain, we may take the limit
$\eta\rightarrow0$ of the $XXZ$ chain, for the $XXZ$ chain with real
$\eta$, we may take a proper limit of $XYZ$ model\cite{cysw3}, and
for the $XXZ$ torus, we may take a proper limit of the $XYZ$ torus.
In fact, for most of the rational integrable models, their
trigonometric and elliptic counterparts exist. The latter ones
normally possess degenerate points and thus the present method
works.
\section*{Acknowledgments}

J. Cao, W.-L. Yang and Y. Wang are grateful for valuable discussions
with R.I. Nepomechie. The financial support from the National
Natural Science Foundation of China (Grant Nos. 11174335,
11031005, 11375141, 11374334), the National Program for Basic
Research of MOST (973 project under grant No.2011CB921700), the
State Education Ministry of China (Grant No. 20116101110017) and
BCMIIS are gratefully acknowledged. Two of the authors (W.-L. Yang
and K. Shi) would like to thank IoP/CAS for the hospitality and they
enjoyed during their visit there.

\end{document}